\documentclass[galaxies,article,accept,pdftex,moreauthors]{Definitions/mdpi}

\usepackage[normalem]{ulem}

\firstpage{1} 
\pubvolume{1}
\issuenum{1}
\articlenumber{0}
\pubyear{2024}
\copyrightyear{2024}
\externaleditor{Academic Editor: Jorick S. Vink}
\datereceived{11 September 2024} 
\daterevised{09 October 2024} % Comment out if no revised date
\dateaccepted{11 October 2024} 
\datepublished{ } 
\hreflink{https://doi.org/} % If needed use \linebreak

\def\ini{\mathrm{ini}}
\def\msun{M_{\odot}}
\def\Msun{$~M_{\odot}$}

\Title{Production of Lithium and Heavy Elements in AGB Stars Experiencing PIEs}

\TitleCitation{Production of Lithium and Heavy Elements in AGB Stars Experiencing PIEs}

\Author{Arthur Choplin $^{1}$*\orcidA{}, Lionel Siess $^{1}$, Stephane Goriely $^{1}$ and Sebastien Martinet $^{1}$}

\AuthorNames{Arthur Choplin, Lionel Siess, Stephane Goriely and Sebastien Martinet}

\AuthorCitation{Choplin, A.; Siess, L.; Goriely, S.; Martinet, S.}
\address{%
$^{1}$ \quad Institut d'Astronomie et d'Astrophysique, Universit\'e Libre de Bruxelles,  CP 226, B-1050 Brussels, Belgium; lionel.siess@ulb.be~(L.S.); stephane.goriely@ulb.be~(S.G.); sebastien.martinet@ulb.be~(S.M.)\\  
}

\corres{Correspondence: arthur.choplin@ulb.be}

\abstract{Asymptotic giant branch (AGB) stars can experience proton ingestion events (PIEs), leading to a rich nucleosynthesis. During a PIE, the intermediate neutron capture process (i-process) develops, leading to the production of trans-iron elements. It is also suggested that lithium is produced during these events. 
We investigate the production of lithium and trans-iron elements in AGB stars experiencing a PIE with $1<M_\ini/\msun < 3$ and $-3< \mathrm{[Fe/H]} <0$. 
We find that lithium is produced in all PIE models with surface abundances $3<$~A(Li)~$<5$. The surface enrichment and overall AGB lithium yield increases with decreasing stellar mass. The lithium enrichment is accompanied by a production of $^{13}$C with $3<^{12}$C/$^{13}$C~$<9$ at the surface just after the PIE. 
AGB stars experiencing PIE may be related to J-type carbon stars whose main features are excesses of lithium and $^{13}$C. 
In addition to Li and $^{13}$C, heavy elements (e.g., Sr, Ba, Eu, Pb) are significantly produced in low-metallicity stars up to [Fe/H]~$\simeq-1$.  
The yields of our models are publicly available. Additionally, of interest to the Li nucleosynthesis, we provide an updated fitting formula for the $^{7}$Be($e^-,\nu_e$)$^{7}$Li electron capture rate.}


\begin{document}

\section{Introduction}

The asymptotic giant branch (AGB) phase corresponds to the life~end of $\sim$1--8 $\msun$  stars (see, e.g., \cite{karakas14} for a review). This stage is characterized by recurrent convective thermal pulses (TPs) that develop in the H-He intershell region. 
When fully developed, the top of the convective pulse nearly reaches the bottom of the H-burning shell. In some cases, protons can be ingested in the TP, leading to a proton ingestion event (PIE, e.g., \cite{fujimoto00, iwamoto04}). When a PIE arises, H atoms are transported downwards in the pulse and burnt on the flight via the $^{12}$C($p,\gamma$)$^{13}$N reaction with $^{13}$N decaying to $^{13}$C in about 10~min. The $^{13}$C($\alpha,n$) reaction also operating at the bottom of the TP, where $T \simeq 250$~MK, leads to neutron densities as high as a few $10^{15}$~cm$^{-3}$, leading to the so-called intermediate neutron capture process (i-process, first coined by \cite{starrfield75, cowan77}) and to the production of trans-iron elements (e.g., \cite{cristallo09a, choplin21}). 
In addition to AGB stars, PIEs (and thus i-process nucleosynthesis) can develop in accreting white dwarfs~\cite{denissenkov17,denissenkov19,piersanti19}, low-metallicity massive stars \cite{pignatari15,banerjee18, clarkson18,clarkson20}, during the core helium flash of low-metallicity low-mass stars \cite{fujimoto90,schlattl01,campbell10,cruz13} and in post-AGB stars experiencing a late thermal pulse \cite{herwig01, herwig11}.

During PIEs, some $^{3}$He is also engulfed in the TP and burnt via $^{3}$He($\alpha,\gamma$)$^{7}$Be. The freshly produced $^{7}$Be is quickly transported by convection to colder regions where $^{7}$Li is produced  by $^{7}$Be($e^-,\nu_e$)$^{7}$Li before being destroyed by $^7$Li($p,\alpha$)$^{4}$He. This $^{7}$Li production channel is known as the Cameron--Fowler Beryllium transport mechanism \cite{cameron71} and has been shown to operate in low-mass low-metallicity AGB stars experiencing PIEs \citep{iwamoto04}.

In this paper, we investigate the production of lithium and heavy elements (and $^{13}$C) in AGB stars experiencing PIEs over a wide range of masses and metallicities.

\section{Computation of AGB Stellar Models}

\subsection{Initial Conditions and Nuclear Network}
The AGB models considered in this work are those computed in Refs.~\cite{choplin22a,choplin24}. Models have initial masses $1<M_\ini/\msun< 3$ and metallicities\endnote{[X/Y]~$= \log_{10}(N_{\rm X} / N_{\rm Y})_{\star} - \log_{10}(N_{\rm X} / N_{\rm Y})_{\odot}$ with $N_{\rm X}$ and $N_{\rm Y}$ the numbers of atoms of elements X and Y in the considered model (or observed star) and in the Sun.}    
$-3<$~[Fe/H]~$<0$ using the solar mixture of \cite{asplund09}. We highlight below some relevant computational aspects and refer the reader to the aforementioned works for additional details. 
The models were computed with the stellar evolution code STAREVOL \cite{siess00, siess06, goriely18c}.
During a PIE, a network of 1160 nuclei is used, comprising 2123 nuclear reactions ($n$-, $p$-, $\alpha$-captures and $\alpha$-decays), weak (electron captures, $\beta$-decays), and electromagnetic interactions of relevance to follow neutron capture processes up to neutron densities of $\sim 10^{17}$~cm$^{-3}$. The nuclear network is coupled to the diffusion equation for the transport of chemical species since, during a PIE, the nuclear burning and convective transport timescales become similar. 
The nominal rate for the electron capture reaction $^{7}$Be($e^-,\nu_e$)$^{7}$Li is from \cite{caughlan88} with a threshold value of \mbox{$\tau_{\rm 1/2} = 53$}~days below $T<10^6$~K. The models presented here were computed with this rate. However, we tested the impact of adopting the theoretical rate of \cite{simonucci13} (also tested in~\cite{vescovi19} to evaluate the impact on the solar neutrino fluxes) and found very little difference in the Li production during the PIE. We provide a new fitting formula to the temperature- and density-dependent rate of \cite{simonucci13} in Appendix~\ref{app:app1}. 

\subsection{Overshoot Mixing}
\label{sect:over}

PIEs have been shown to develop (even without extra mixing) during the early AGB phase of stars with $M_\ini < 2.5~\msun$  and [Fe/H]~$<-2$ \cite{iwamoto04, campbell08, cristallo09a, cristallo16, lau09, suda10, gilpons22, choplin21, choplin22a, choplin24, remple24}, considering  overshooting above the convective TP facilitates PIEs \cite{choplin24}. Depending on the strength of overshoot (which is poorly constrained), PIEs can develop up to $M_\ini = 4~\msun$  and [Fe/H]~$=0$ (\cite{choplin24}; \cite{choplin24b}). 
In some of the models considered here, overshooting above the TP is included following the exponential law introduced in \cite{goriely18c}. For these models, the depth of the mixing beyond the Schwarzschild boundary is controlled by the parameter $f_{\rm top}$ \citep[cf. Sect.~2 in][]{choplin24}, the value of which is set to $f_{\rm top} = 0.02$, $0.04$ or $0.10$. Setting $f_{\rm top}=0.10$ triggers a PIE in all our models except at solar metallicity ([Fe/H]~$=0$). However, a 2\Msun\ model at [Fe/H]~$=0$ with $f_{\rm top} = 0.12$ can develop a PIE.

%%%%%%%%%%%%%%%%%%%%%%%%%%%%%%%%%%%%%%%%%%
\section{Results}

\subsection{Lithium Production in a 1\Msun\ [Fe/H]~$=-2.5$ AGB Model }
\label{sect:liprod}

\begin{figure}[H]

\includegraphics[width=0.9\columnwidth]{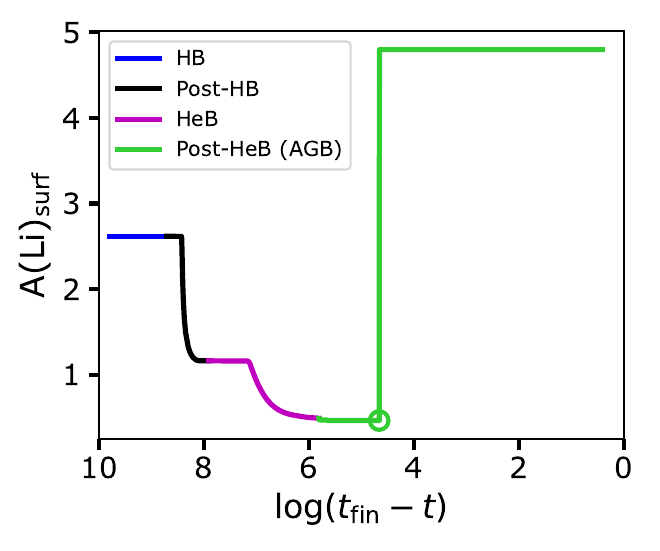}
\caption{Surface evolution of Li abundance in a 1\Msun, [Fe/H]~$=-2.5$ asymptotic giant branch (AGB) star model. 
The different colors represent the different evolutionary stages (HB for core hydrogen burning, HeB for core helium burning). The circle denotes the occurrence of a proton ingestion event (PIE).}
\label{fig:alisurf}
\end{figure}

\begin{figure}[H]

\includegraphics[width=1\columnwidth]{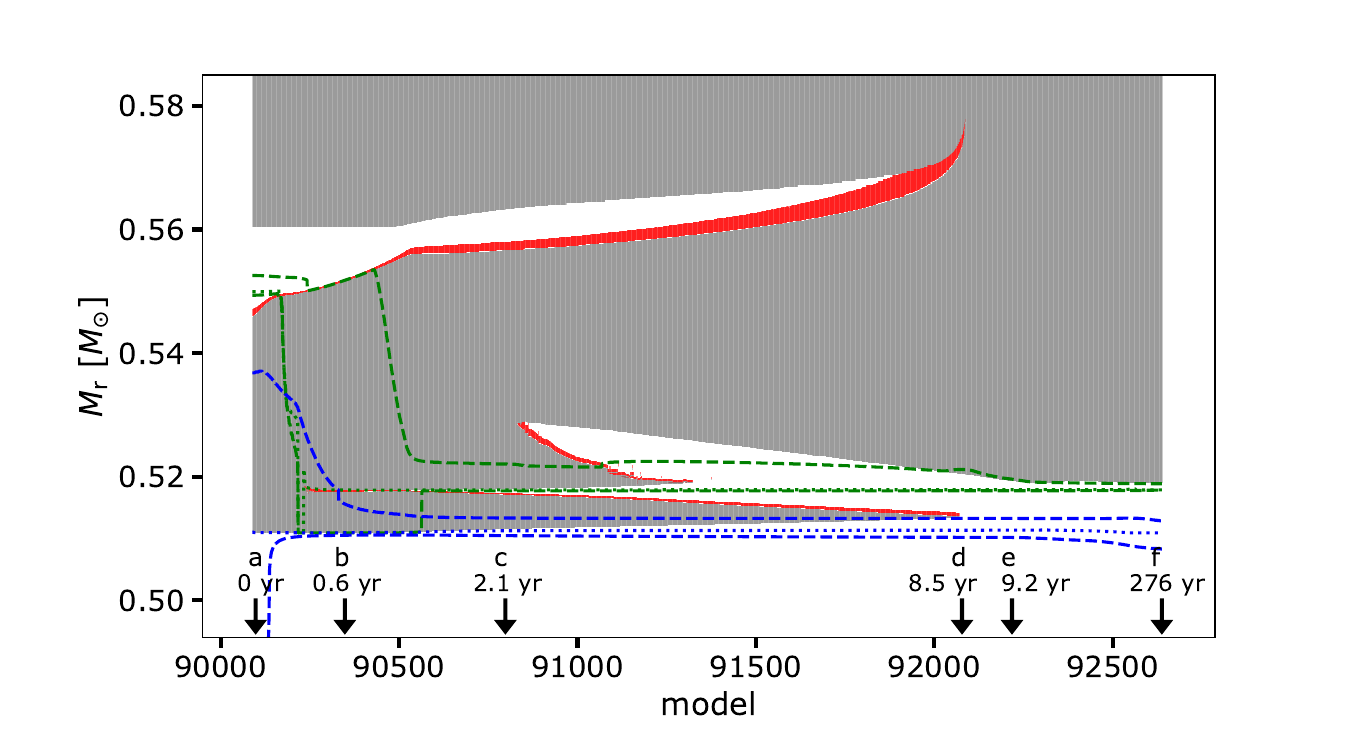}
\includegraphics[width=1\columnwidth]{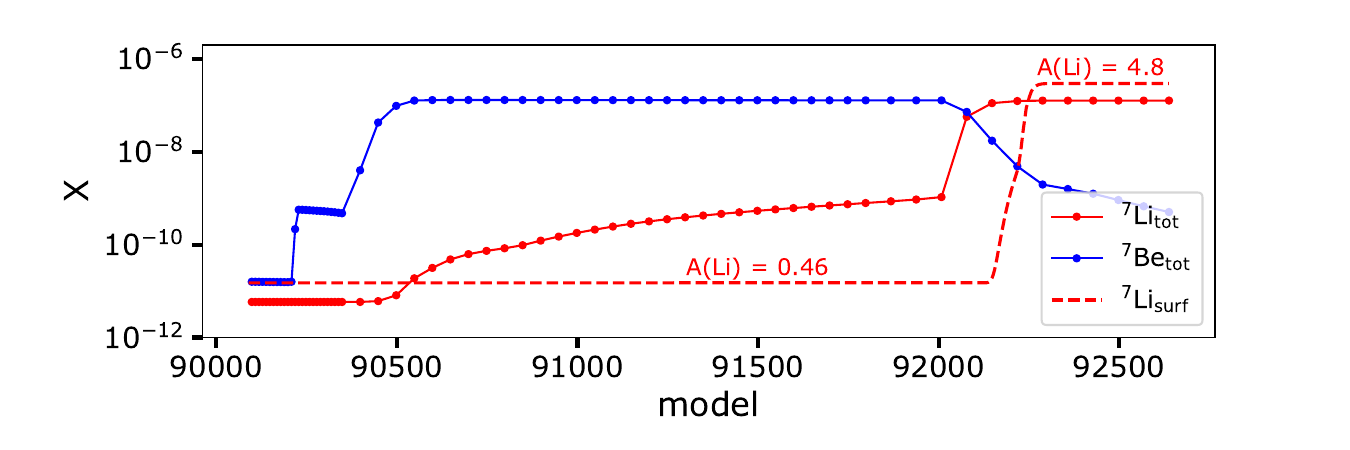}
\caption{Top: Kippenhahn diagram showing the PIE in a 1~$M_{\odot}$, [Fe/H]~$=-2.5$ AGB model.
Convective regions are shaded gray. 
The thermal pulse has already developed at point a (0 yr).
The dotted green and blue lines trace the mass coordinate where the nuclear energy production by hydrogen and helium burning is maximum, respectively. 
The dashed green and blue lines delineate the hydrogen and helium-burning zones (where the nuclear energy production by H and He burning exceeds 10~erg~g$^{-1}$~s$^{-1}$). The red areas show the extent of the overshoot zones.
The letters at the bottom correspond to the abundance profiles shown in Figure~\ref{fig:abund}. The numbers indicate the time in years since the first model (90098). 
Bottom: evolution of the average $^{7}$Li (solid red) and $^{7}$Be (solid blue) stellar mass fractions (cf. text for details). The red dashed line shows the surface $^{7}$Li mass fraction (and corresponding abundance A(Li)).
}
\label{fig:kip}
\end{figure}

Lithium is a fragile element that is destroyed by proton captures when the temperature exceeds $\sim 2.5$~MK. On the one hand, Li is burnt during the pre-main sequence phase, and its abundance further diluted during the first dredge-up event.
In our 1\Msun\ [Fe/H]~$=-2.5$ AGB model, the surface Li abundance\endnote{A(Li)~$= \log_{10} (N_{\rm Li} / N_{\rm H}) + 12$, where $N_{\rm Li}$ and $N_{\rm H}$ refer to the numbers of atoms of lithium and hydrogen, respectively. A(Li)} drops to about 1 after the first dredge up (Figure~\ref{fig:alisurf}, at $\log \, $($t_{\rm fin} - t$)~$\simeq 8$). 
The A(Li) value further decreases to $0.5$ during the early AGB phase, when the convective envelope reaches deep layers (at $\log \, $($t_{\rm fin} - t$)~$\simeq 6$). 
During the PIE (shown in Figure~\ref{fig:kip} top panel), the average\endnote{At a given time, the average stellar mass fraction of species $i$  is given by $\overline{X_{\rm i}} = (1 / M_{\rm tot}) \int_{0}^{M_{\rm tot}} X_{\rm i} (M_{\rm r}) \, \mathrm{d} m$ with $M_{\rm tot}$ the stellar mass and $X_{\rm i} (M_{\rm r})$ the mass fraction of chemical species $i$ at mass coordinate $M_{\rm r}$.} stellar $^{7}$Be and  $^{7}$Li mass fraction dramatically increase (Figure~\ref{fig:kip} bottom panel), reaching A(Li) values after the PIE up to 4.8 (Figure~\ref{fig:alisurf}).

\begin{figure}[H]

 \begin{minipage}[c]{1\columnwidth}
\includegraphics[width=0.5\columnwidth]{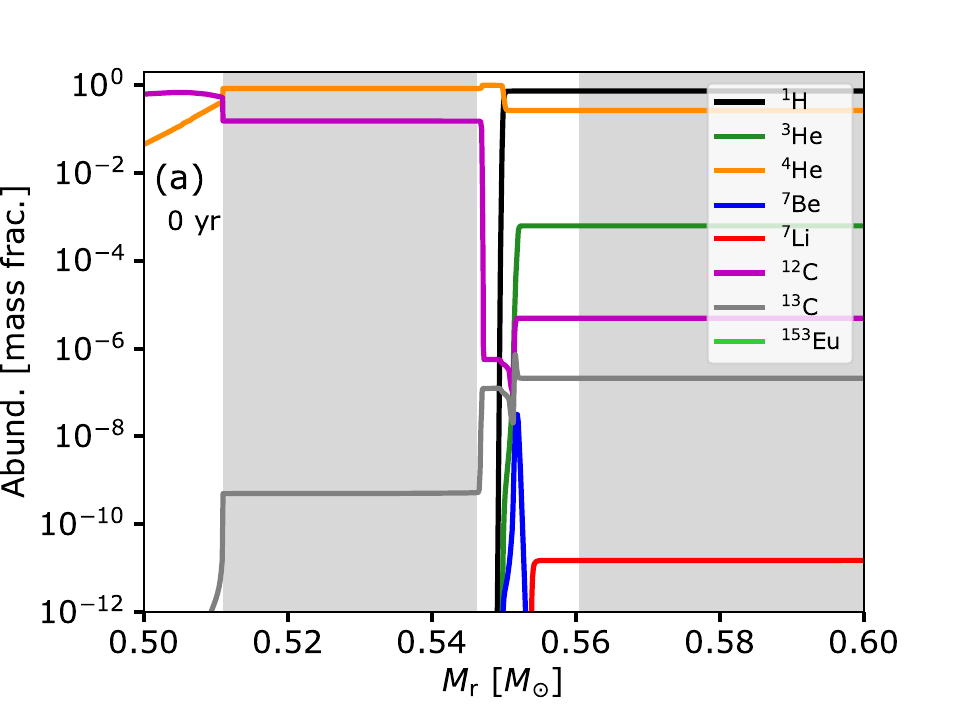}
\includegraphics[width=0.5\columnwidth]{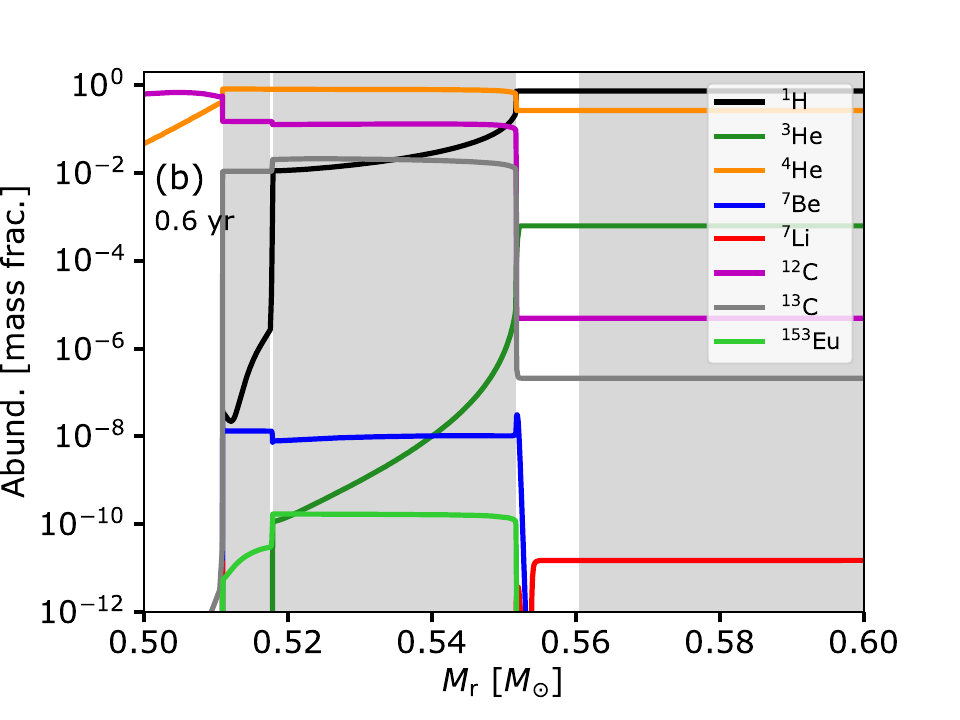}
  \end{minipage}
 \begin{minipage}[c]{1\columnwidth}
\includegraphics[width=0.5\columnwidth]{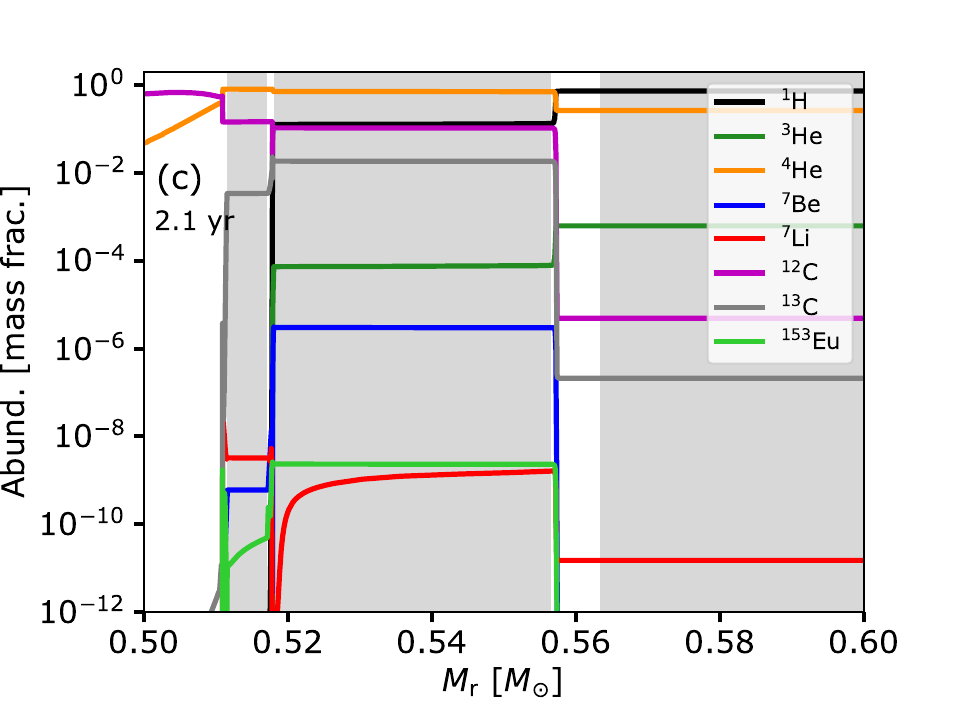}
\includegraphics[width=0.5\columnwidth]{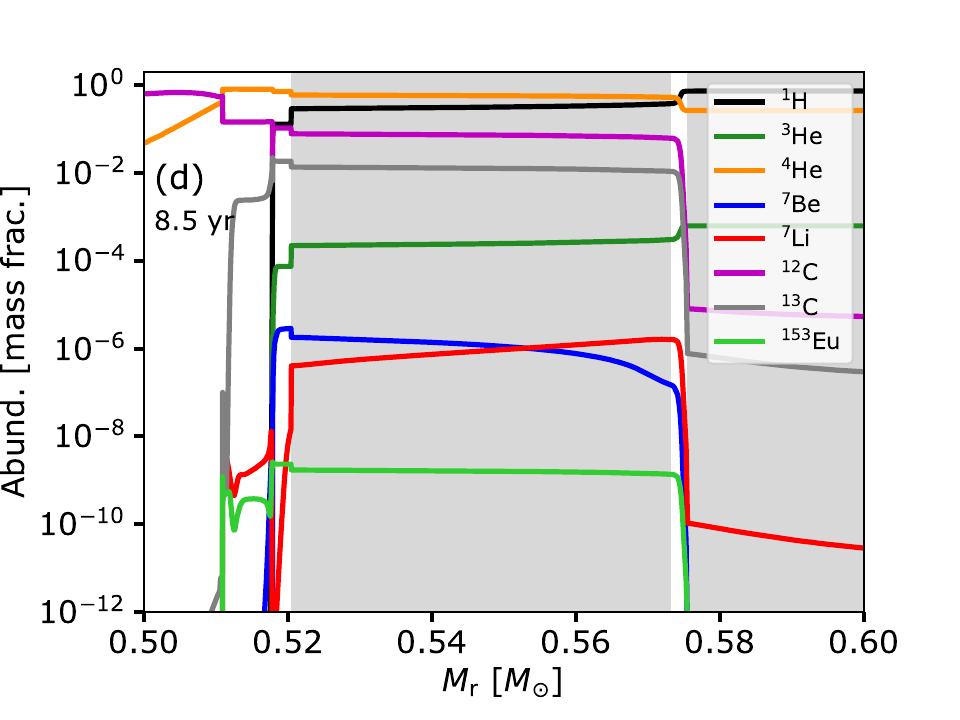}
  \end{minipage}
 \begin{minipage}[c]{1\columnwidth}
\includegraphics[width=0.5\columnwidth]{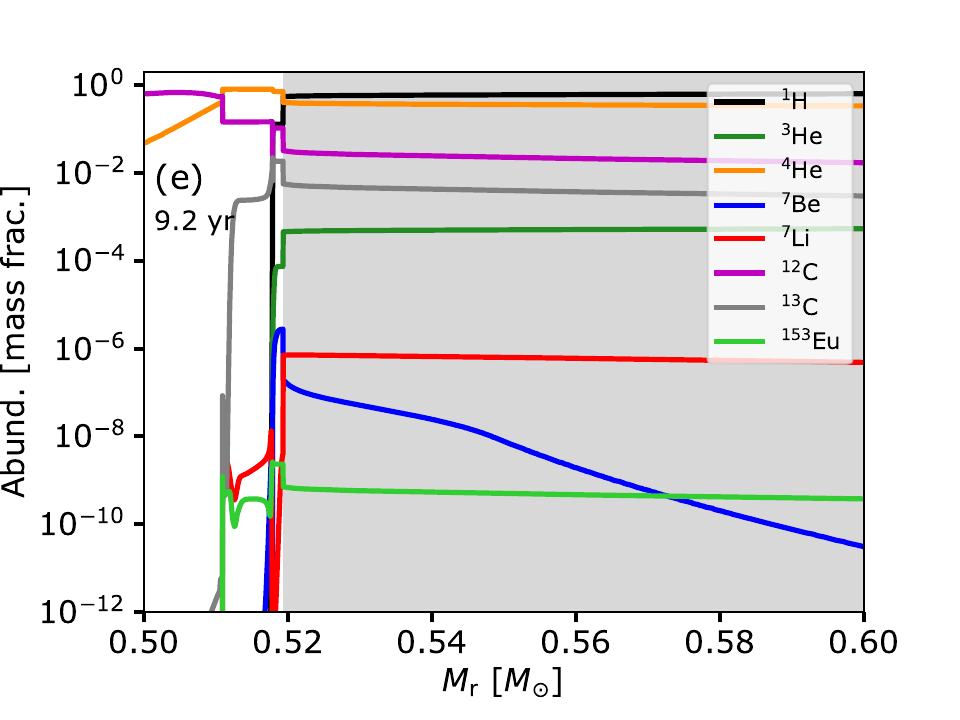}
\includegraphics[width=0.5\columnwidth]{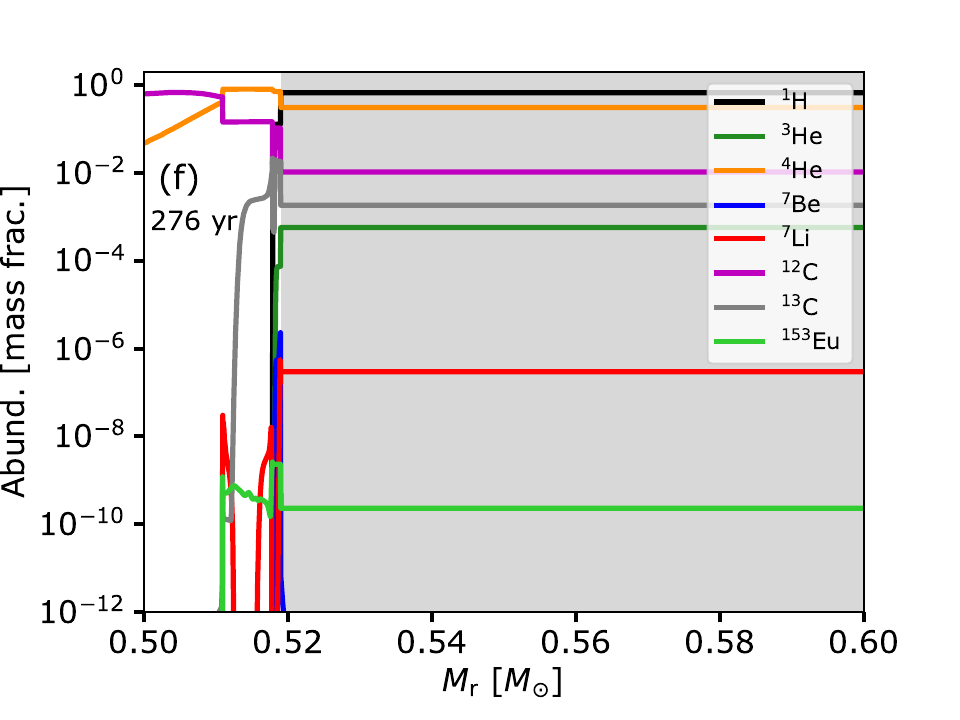}
  \end{minipage}
\caption{Abundance profiles in the inner layers of a 1~$M_{\odot}$, [Fe/H]~$=-2.5$ AGB model at $t=0$~yr (a), $t=0.6$~yr (b), $t=2.1$~yr (c), $t=8.5$~yr (d), $t=9.2$~yr (e) and $t=276$~yr (f). 
The different panels correspond to the letters indicated in Figure~\ref{fig:kip} top panel.
The time since the first model (top~left) is indicated.
The gray area show the convective zones (corresponding to the the gray zones of~Figure~\ref{fig:kip}.}
\label{fig:abund}
\end{figure}

We describe below in detail the lithium production in this model. 
The abundance profiles just before the PIE are shown in Figure~\ref{fig:abund}a. In the outer layers, the mass fraction of $^{3}$He is about $10^{-3}$. Some $^{7}$Be is present in the top of the H burning shell at $M_{\rm r} \sim 0.55~\msun$  because of the operation of the $^{3}$He($\alpha,\gamma$)$^{7}$Be reaction. 
As the PIE develops, protons and $^{3}$He are engulfed in the convective pulse (Figure~\ref{fig:abund}b) and $^{7}$Be is further produced.
After about 2~yr, the mass fraction of $^{7}$Be in the pulse has increased to $\simeq 10^{-6}$ (Figure~\ref{fig:abund}c). Some $^{7}$Li (in red) starts to be synthesized by $^{7}$Be($e^-,\nu_e$)$^{7}$Li.
As the convective pulse grows and engulfs more $^{3}$He, the average $^{7}$Be mass fraction in the star increases (bottom panel of Figure~\ref{fig:kip}). Electron captures on $^{7}$Be progressively raise the overall abundance of $^{7}$Li in the star (red solid line in the bottom panel of Figure~\ref{fig:kip}). From model $\sim 90500$, no more $^{7}$Be is synthesized  despite the fact that the growing pulse is engulfing $^{3}$He (see the plateau in the bottom panel of Figure~\ref{fig:kip}) because the temperature in the pulse is too low to activate $^{3}$He($\alpha,\gamma$)$^{7}$Be. 
After $8.5$~yr (Figure~\ref{fig:abund}d), the pulse is about to merge with the envelope, and $^{7}$Li has been significantly synthesized. After the merging, $^{7}$Be and $^{7}$Li quickly reach the surface (Figure~\ref{fig:abund}e) where the abundance of A(Li) raises from 0.46 to 4.8 (Figure~\ref{fig:kip}, bottom panel).
After a few decades, the $^{7}$Be in the envelope has been fully transformed into $^{7}$Li (Figure~\ref{fig:abund}f).

\vspace{-6pt}

\subsection{PIE Lithium Production as a Function of Mass and Metallicity}
\label{sect:agbmz}

\begin{figure}[H]

\includegraphics[width=0.9\columnwidth]{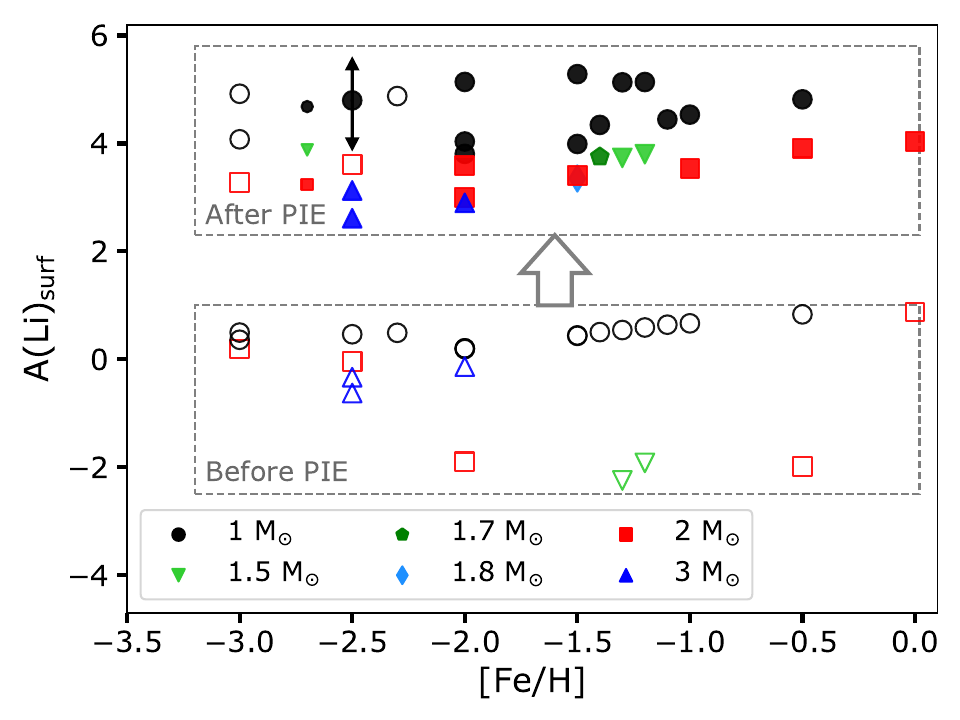}
\caption{Surface Li abundance just before (lower rectangle) and just after the PIE (upper rectangle) in AGB models of various masses and metallicities. 
Different colors correspond to different initial masses. 
In the upper rectangle, empty (filled) symbols correspond to models computed without (with) overshoot. Some overshoot models have same mass and metallicities but were computed with different $f_{\rm top}$ values (cf. Section~\ref{sect:over}). 
The small upward (downward) black arrow shows how the surface A(Li) is impacted in a 1~$M_{\odot}$, [Fe/H]~$=-2.5$ model if considering 10 times more (less) $^{3}$He in the envelope just before the PIE.
 The three smaller symbols at [Fe/H]~$=-2.7$ correspond to the surface A(Li) of the 1, 1.5 and 2\Msun\ models  just after the PIE, computed by \cite{iwamoto04}.
}
\label{fig:libefaft}
\end{figure}

All our PIE models experience a similar lithium evolution as described in Section~\ref{sect:liprod} (with an exception for 3\Msun\ models as discussed later): dilution during the first and second dredge ups and then production during the PIE. At the beginning of the AGB phase, the surface A(Li) ranges from $-2.2$ to $0.9$ (Figure~\ref{fig:libefaft}, symbols in the lower rectangle) and reaches values of $2.6 <$~A(Li)~$< 5.3$ after the PIE (symbols in the upper rectangle).  
This Li production is sensitive to (1) the amount of $^{3}$He in the envelope right before the PIE, (2)~the PIE characteristics (especially the thermodynamic conditions) and (3) the size of the convective pulse compared to the envelope. These three aspects are discussed below in more~detail.

\begin{figure}[H]
 
\includegraphics[width=0.9\columnwidth]{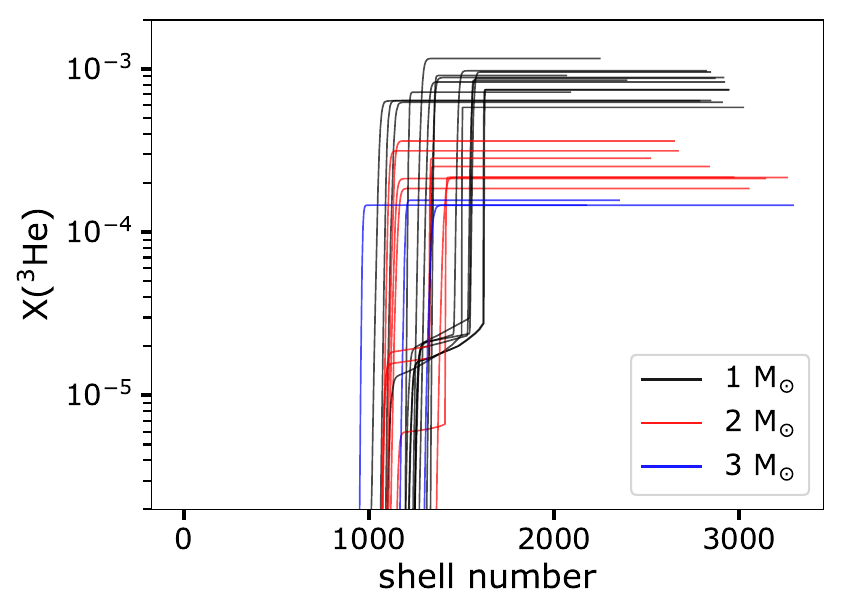}
\caption{$^{3}$He abundance profile in our 1, 2 and 3\Msun\ AGB models just before the PIE. 
}
\label{fig:he3li7p}
\end{figure}

\newpage
\begin{enumerate}
\item $^{3}$He is massively synthesized in low mass stars during core H-burning and  \cite{rood76} showed a positive correlation between the main sequence duration, i.e., the initial stellar mass and  the production of $^{3}$He. Following the first dredge up that brings $^{3}$He to the surface, the following evolutionary stages weakly affect its  abundance, provided extra mixing mechanisms like thermohaline or rotation are not included (e.g., \cite{charbonnel95, charbonnel07, eggleton08, lagarde11, lagarde12}). 
Given the higher $^{3}$He abundance in the lower mass stars at the time of the PIE (Figure~\ref{fig:he3li7p}), we would expect a higher production of Li in these lower mass models. As a numerical test, we increased and decreased the $^{3}$He mass fraction by a factor of 10 in the envelope of a 1\Msun\ at [Fe/H]~$=-2.5$ right before the PIE. The resulting surface A(Li) after the PIE is accordingly impacted by a factor of $\sim$10 (black arrows in Figure~\ref{fig:libefaft}). 

\item As the PIE proceeds, the layers of the pulse expand and the temperature decreases. The production of $^{7}$Be (through $^{3}$He($\alpha,\gamma$)$^{7}$Be) thus depends on how long a sufficiently high temperature is maintained in the convective pulse. This is determined by the pulse characteristics which vary with initial mass, metallicity and, in a given model, from pulse to pulse. 
During a PIE, our high-mass AGB models maintain a large temperature longer at the bottom of the pulse before the merging of the pulse with the envelope.  
This tends to favor the production of $^{7}$Li in higher mass stars. 

\item Finally, with increasing initial mass, the extent of the pulse decreases while the mass of the envelope increases. For example, in our 
1, 2 and 3 $M_{\odot}$ models at [Fe/H]~$=-2$, the ratio of the envelope to  pulse mass is $\approx 4$, 40 and 200, respectively. 
The lithium is thus more diluted in the more massive envelope of higher mass stars, which results in smaller Li surface enrichments. 

\end{enumerate}

\begin{figure}[H] 
\includegraphics[width=0.9\columnwidth]{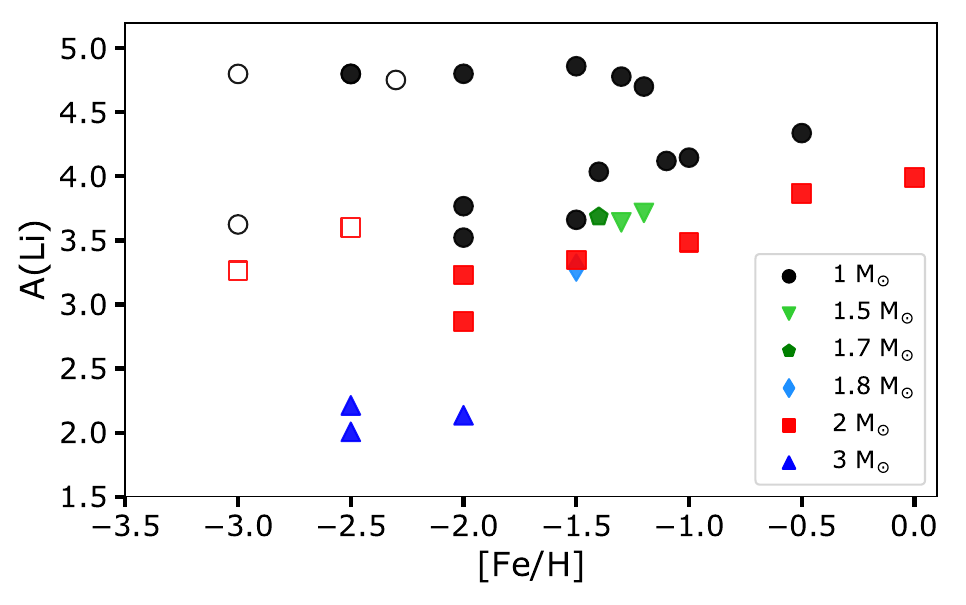}
\caption{A(Li) in the yields of our $1-3~\msun$  AGB models experiencing a PIE. 
Different colors correspond to different initial masses. 
Empty (filled) symbols correspond to models computed without (with) overshoot. Some overshoot models have same mass and metallicities but were computed with different $f_{\rm top}$ values (cf. Section~\ref{sect:over}).
}
\label{fig:aliyie}
\end{figure}

To summarize, (1) and (3) tend to favor higher Li enrichment in lower mass stars, while (2) leads to higher Li enrichment in higher mass stars. 
All together, these effects lead to higher Li enrichment in lower mass stars (Figure~\ref{fig:libefaft} and \ref{fig:aliyie}). 
Right after the PIE, $3\lesssim$~A(Li)~$\lesssim5$ at the surface (Figure~\ref{fig:libefaft}) with a rather clear dependence on the initial mass.
After the PIE, 1\Msun\ models quickly lose their (Li-rich) envelope without experiencing any further thermal pulse. In contrast, the AGB phase resumes in more massive models. Lithium in the envelope is however left unchanged until the end of the AGB except in the 3\Msun\ models, where it is partly burnt due to the high temperature ($T>2.5$~MK) at the bottom of the convective envelope.
As a consequence, the Li abundances A(Li), corresponding to our AGB model yields\endnote{The yield $\mathcal{Y}_i$ of a nucleus $i$ is computed according to the relation $\mathcal{Y}_i =  \int_{0}^{\tau_{\rm star}} \dot{M}(t) \, X_{\rm i,s}(t) \, \text{d}t$ where $\tau_{\rm star}$ is the total lifetime of the model star, and $X_{\rm i,s}(t)$ and $\dot{M}(t)$ are the surface mass fraction of the nucleus, $i,$ and the mass-loss rate at time, $t$, respectively. The A(Li) value in the yields can then be computed as A(Li)~$= \log_{10} (N_{\rm Li} / N_{\rm H}) + 12 = \log_{10} [(\mathcal{Y}_{\rm ^{6}Li} / 6 +  \mathcal{Y}_{\rm ^{7}Li} / 7) / \mathcal{Y}_{\rm ^{1}H}] + 12$.} (Figure~\ref{fig:aliyie}), are similar to the surface Li abundance after the PIE, except for the 3\Msun\ models where lithium is partially destroyed in the envelope, so that the ejected material becomes Li-poor.
We note that the surface A(Li) values of the three PIE AGB models computed in \cite{iwamoto04} (small symbols in Figure~\ref{fig:libefaft}) are in agreement with our~models.

\vspace{-6pt}

\vspace{-6pt}

\subsection{Production of $^{13}$C}

\begin{figure}[H]

\includegraphics[width=0.9\columnwidth]{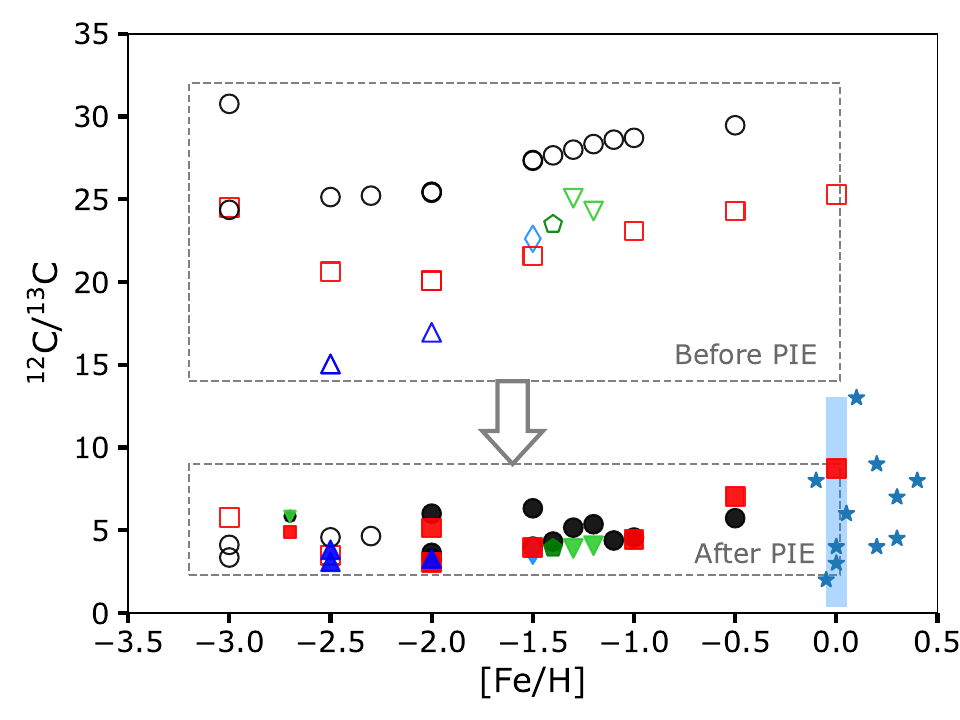}
\caption{Surface $^{12}$C/$^{13}$C ratios just before (upper rectangle) and just after the PIE (lower rectangle) in AGB models of various masses and metallicities. 
Different colors correspond to different initial masses (as in Fig.~\ref{fig:libefaft} and \ref{fig:aliyie}). 
In the lower rectangle, empty (filled) symbols correspond to models computed without (with) overshoot. Some overshoot models have the same mass and metallicities but were computed with different $f_{\rm top}$ values (cf. Section~\ref{sect:over}). 
The three smaller symbols at [Fe/H]~$=-2.7$ show the surface $^{12}$C/$^{13}$C ratio of the 1, 1.5 and 2\Msun\ models of \cite{iwamoto04}, just after the PIE. 
The blue stars represent J-type stars \citep{abia97, abia00, hedrosa13, abia17} and the blue shaded area show the range of $^{12}$C/$^{13}$C ratios in  a sample of post-AGB sources (Table~2 in \cite{ziurys20}).
}
\label{fig:ccbefaft}
\end{figure}

During the first dredge up, the ashes of H burning and among them $^{13}$C are transported to the stellar surface. 
In standard models, this leads to a decrease in the surface $^{12}$C/$^{13}$C ratio from 89 (solar value) to $\sim$15--30  (e.g., \cite{charbonnel94}). 
Without extra-mixing processes induced by, e.g., rotation, the $^{12}$C/$^{13}$C ratio in low-mass stars is not expected to change significantly until the early AGB. 
Subsequently, recurrent third dredge-ups will inject $^{12}$C-rich pulse material into the envelope and the ratio will increase.
In our standard solar metallicity 2\Msun\ model that does not experience a PIE, $^{12}$C/$^{13}$C drops to 26 after the first dredge up and progressively rises during the AGB phase to reach a final value of 81.  
Including in this model strong overshooting above the pulse triggers a PIE, during which $^{13}$C is synthesized by $^{12}$C($p,\gamma$)$^{13}$N($\beta^+$)$^{13}$C. The $^{13}$C isotope is later dredged up to the surface when the pulse merges with the envelope, decreasing the $^{12}$C/$^{13}$C ratio from 26 to 9. Considering all our models, the surface $^{12}$C/$^{13}$C ratio decreases from 15--31 just before the PIE to 3--9 just after (Figure~\ref{fig:ccbefaft}), i.e.,  close to the CNO equilibrium value of about 4.
The models of \cite{iwamoto04} at [Fe/H]~$=-2.7$ predict a similar low surface $^{12}$C/$^{13}$C ratio right after the PIE (small symbols in Figure~\ref{fig:ccbefaft}). 
We note that in  AGB stars with $M_\ini \gtrsim 1.5 ~\msun$, additional pulses and 3DUPs develop after the PIE \cite{choplin22a, choplin24}, potentially increasing the $^{12}$C/$^{13}$C ratio (\cite{choplin24b}).

\subsection{Heavy Element Production with the i-Process}
\label{sect:ipro}

The production of heavy elements by the i-process in PIE was extensively discussed in, e.g., \cite{cristallo09a, choplin21, choplin22a, choplin22b, choplin24, martinet24}. We discuss here a few aspects related to the models presented in~\cite{choplin22a, choplin24} and refer to these works for more details.

\begin{figure}[H]
\centering
 \begin{minipage}[c]{1\columnwidth}
\includegraphics[width=0.5\columnwidth]{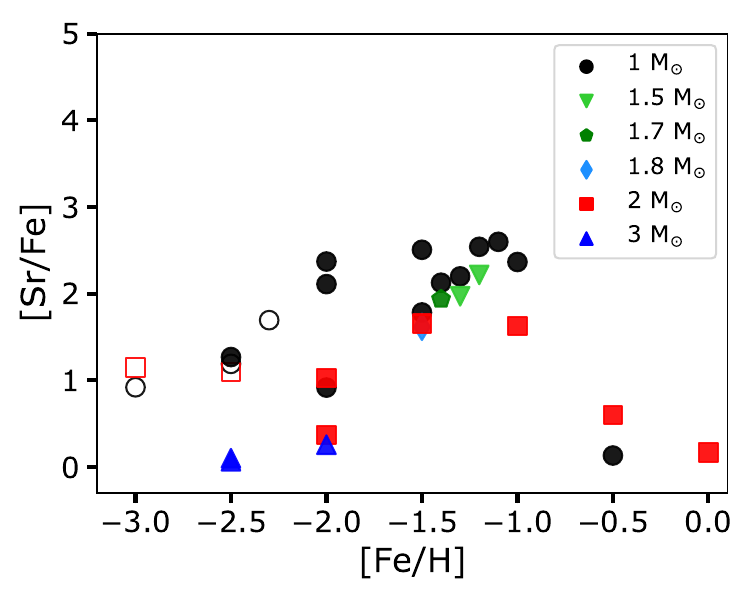}
\includegraphics[width=0.5\columnwidth]{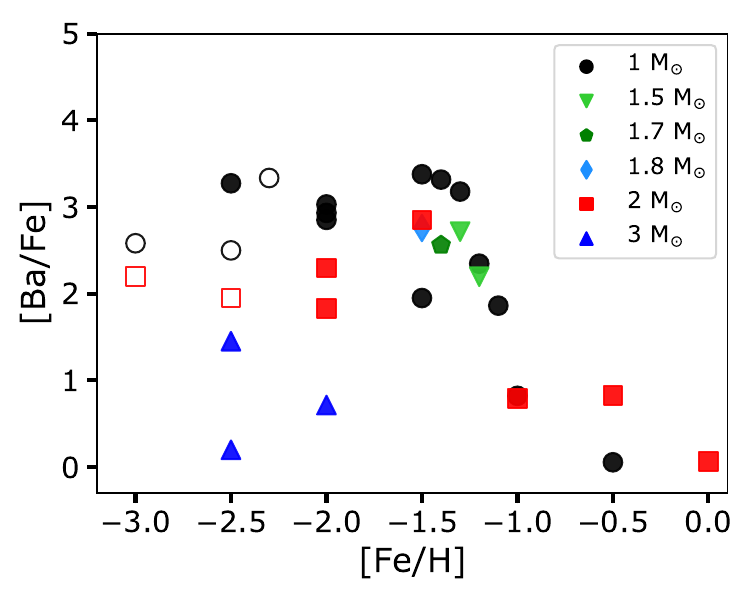}
  \end{minipage}
 \begin{minipage}[c]{1\columnwidth}
\includegraphics[width=0.5\columnwidth]{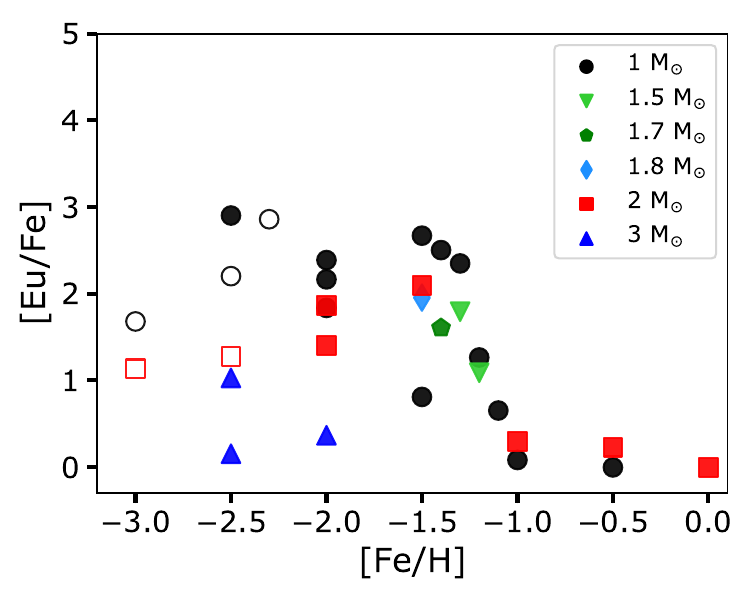}
\includegraphics[width=0.5\columnwidth]{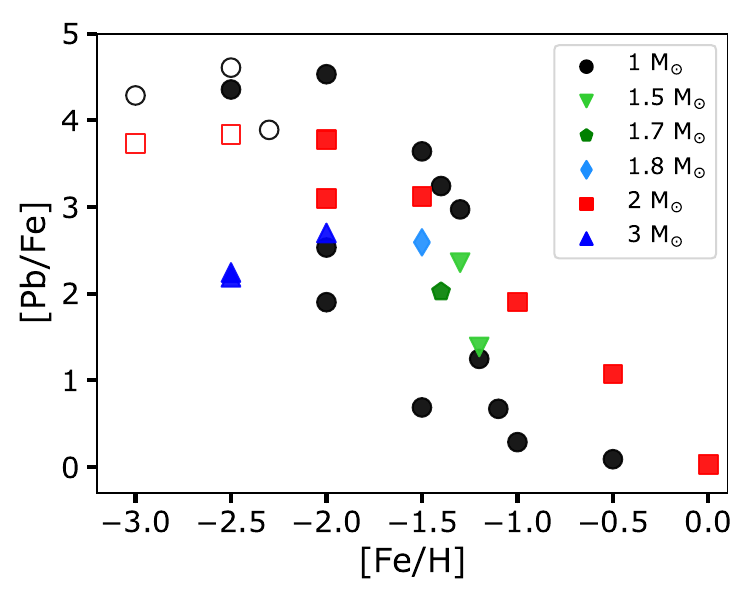}
  \end{minipage}
\caption{[Sr/Fe], [Ba/Fe], [Eu/Fe] and [Pb/Fe] ratios as a function of metallicity in the yields of our AGB models that experienced a PIE. Different colors correspond to different initial masses (as in Fig.~\ref{fig:libefaft}, \ref{fig:aliyie} and \ref{fig:ccbefaft}). 
Empty (filled) symbols correspond to models computed without (with) overshoot. 
}
\label{fig:iproyie}
\end{figure}

\textls[-25]{The maximal neutron density during the PIE shows no clear dependencies on initial mass and metallicity. It varies between $6.8 \times 10^{13}$~cm$^{-3}$ and $2.2 \times 10^{15}$~cm$^{-3}$. 
\mbox{At [Fe/H]~$\geq-0.5$}}, the production of heavy elements is very small (Figure~\ref{fig:iproyie}): the amount of seed (mainly $^{56}$Fe) increases with metallicity, leading to smaller neutron-to-seed ratios. This prevents a significant production of heavy elements at high metallicities. 
The production of Sr is maximal at $-2<$~[Fe/H]~$<-1$ (Figure~\ref{fig:iproyie}, top left panel). At [Fe/H]~$<-2$, the neutron-to-seed ratio is high enough for (some) Sr to be processed in heavier elements (especially Pb). 
The production of Ba and Eu shows a clear drop between [Fe/H]~$-1.5$ and $-1$. They are consistently produced below [Fe/H]~$=-1.5$, especially by the 1 and 2\Msun\ models.
The Pb yields increase with decreasing metallicity due to the higher neutron-to-seed ratio. 
Often, at a given metallicity, higher mass stars show smaller surface enrichments (Figure~\ref{fig:iproyie}) as a consequence of larger dilution factors. 
Finally, as for lithium (cf. Section~\ref{sect:agbmz}), overshooting can impact the yields. A strong overshooting can, for instance, trigger a PIE in an earlier TP, hence leading to different physical conditions and ultimately to a (generally slightly) different nucleosynthesis (e.g., the two black points at [Fe/H]~$=-2.5$, Figure~\ref{fig:iproyie}).

\section{Conclusions}

AGB stars can experience a proton ingestion event during the early AGB phase. 
This mechanism leads to the intermediate neutron capture process, which significantly produces heavy elements up to [Fe/H]~$\simeq -1$. The production is particularly high in our $M_\ini \simeq$ 1--2~$\msun$  AGB models. 
PIEs also synthesize significant $^{7}$Li, coming from the chain $^{3}$He($\alpha,\gamma$)$^{7}$Be($e^-,\nu_e$)$^{7}$Li. In our models, spanning the mass range $1<M_\ini/\msun< 3$ and metallicity range $-3<$~[Fe/H]~$<0$, a surface enrichment of $3<$~A(Li)~$<5$ is noticed just after the PIE. 
This lithium enrichment comes with an enrichment in $^{13}$C (also synthesized during the PIE), leading to surface $^{12}$C/$^{13}$C ratios of $\sim 3$ to $\sim 9$ depending on the model. 
AGB models with $M_\ini \gtrsim 1.5~\msun$  experience further thermal pulses and third dredge-up episodes after the PIE. 
This destroys some lithium in 3\Msun\ models and tends to increase (slightly) the surface $^{12}$C/$^{13}$C ratio. Additional mixing mechanisms like thermohaline or rotation (not considered here) may change the picture. 
Interestingly, there exists a class of carbon stars called J-type, whose origin is still debated (e.g., \cite{abia20}) and whose main characteristics are the presence of lithium and $^{13}$C. As recently shown in \cite{choplin24b}, most of the properties of J-type stars can be reproduced by AGB models experiencing a PIE, in particular their $^{12}$C/$^{13}$C ratio (blue stars in Figure~\ref{fig:ccbefaft}). 
Interestingly, some post-AGB sources also have very low $^{12}$C/$^{13}$C ratios (blue shaded area in Figure~\ref{fig:ccbefaft} and \citep{schmidt18, ziurys20}).  
They might be the descendants of C-rich stars that left the AGB phase shortly after a PIE.
Further investigations are required to determine if this scenario is viable.

\newpage
\vspace{6pt}

%%%%%%%%%%%%%%%%%%%%%%%%%%%%%%%%%%%%%%%%%%
\authorcontributions{Conceptualization, A.C.; methodology, A.C.; software, L.S., S.G.; validation, L.S., S.G. and S.M.; formal analysis, A.C.; writing---original draft preparation, A.C.; writing---review and editing, L.S., S.G. and S.M. All authors have read and agreed to the published version of the manuscript.} 
%MDPI: For research articles with several authors, a short paragraph specifying their individual contributions must be provided. The following statements should be used ``Conceptualization, X.X. and Y.Y.; methodology, X.X.; software, X.X.; validation, X.X., Y.Y. and Z.Z.; formal analysis, X.X.; investigation, X.X.; resources, X.X.; data curation, X.X.; writing---original draft preparation, X.X.; writing---review and editing, X.X.; visualization, X.X.; supervision, X.X.; project administration, X.X.; funding acquisition, Y.Y. All authors have read and agreed to the published version of the manuscript.'', please turn to the  \href{http://img.mdpi.org/data/contributor-role-instruction.pdf}{CRediT taxonomy} for the term explanation. Authorship must be limited to those who have contributed substantially to the work~reported.}}

\funding{This work was supported by the European Union (ChECTEC-INFRA, project no.~101008324).}
%{Please add: ``This research received no external funding'' or ``This research was funded by NAME OF FUNDER grant number XXX.'' and  and ``The APC was funded by XXX''. Check carefully that the details given are accurate and use the standard spelling of funding agency names at \url{https://search.crossref.org/funding}, any errors may affect your future funding.}}

\dataavailability{The chemical yields of the models presented in this work are publicly available at \url{http://www.astro.ulb.ac.be/~siess/StellarModels/PIE} (accessed on 14 October 2024).  }

\acknowledgments{A.C. is a F.R.S-FNRS fellow. L.S. and S.G. are F.R.S-FNRS research associates. The authors are members of BLU-ULB, the interfaculty research group focusing on space research at ULB - Université libre de Bruxelles. }

\conflictsofinterest{The authors declare no conflicts of interest.}

%%%%%%%%%%%%%%%%%%%%%%%%%%%%%%%%%%%%%%%%%%
%% Optional
\appendixtitles{yes} % Leave argument "no" if all appendix headings stay EMPTY (then no dot is printed after "Appendix A"). If the appendix sections contain a heading then change the argument to "yes".
\appendixstart
\appendix
\section[\appendixname~\thesection]{Analytical Fit to the $^{7}$Be + e$^-$ $\rightarrow$ $^{7}$Li + $\nu_e$ Decay Rate}
\label{app:app1}

\begin{figure}[H]

\includegraphics[width=0.9\columnwidth]{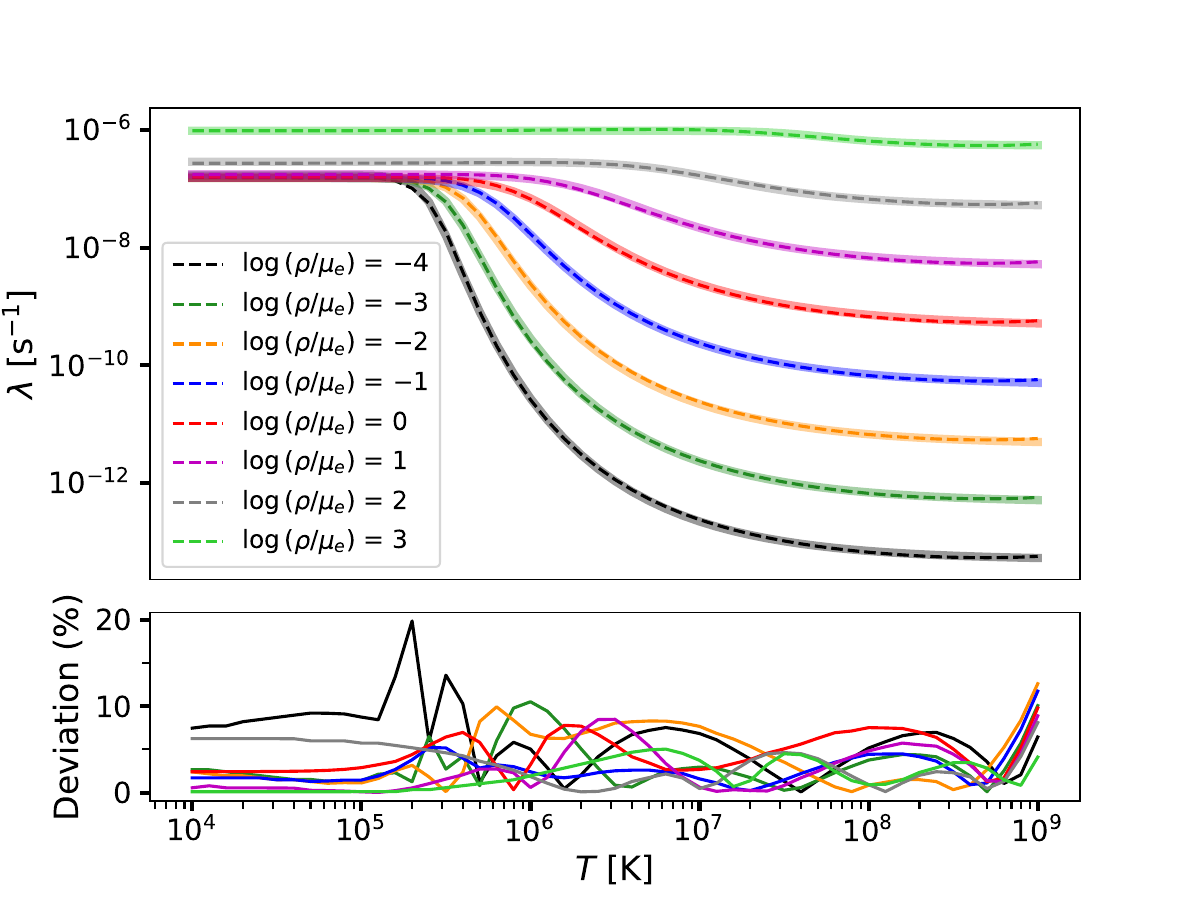}
\caption{Top panel: $^{7}$Be($e^-,\nu_e$)$^{7}$Li rate for various $\rho/\mu_e$ values. Dashed lines correspond to the tabulated rate of \cite{simonucci13, vescovi19} and solid lines show our fits according to Equation~\eqref{eq:fit}.
Bottom panel: deviation between the fit and the tabulated rate in \%.
}
\label{fig:fitbe1}   
\end{figure}

The electron-capture rate of $^{7}$Be strongly depends on the density and temperature. We fitted the theoretical rate $\lambda$ of \cite{simonucci13} starting from the tabulated form given in Ref.~ \cite{vescovi19}. The rate is a function of $\rho/\mu_e$, the density divided by the electron mean molecular weight (\mbox{$-4.1<\log_{10}(\rho/\mu_e)<3$}), and temperature ($10^4<T(\mathrm{K})<10^9$). 
We adopted the  fitting formula 
\begin{linenomath}
\begin{equation}
\log_{10}(\lambda) = \frac{a_0}{(\exp((x-a_1)/a_2) + a_3)^{a_4}}  + a_5
\label{eq:fit}
\end{equation}
\end{linenomath}
with $x=\log_{10}(T)$ and where the coefficients $a_k,\ k=1,5$ are polynomial functions of  $y=\log_{10}(\rho/\mu_e$)
\begin{linenomath}
\begin{equation}
a_k = \sum_{i=0}^{n_k}\,b_{k,i}\,y^{i}
\end{equation}
\end{linenomath}
where $n_k$ is the polynom order ($n_0=4$, $n_1=3$, $n_2=4$, $n_3=1$, $n_4=4$, $n_5=2$) and the parameters $b_{k,i}$ are given in Table~\ref{table:fits}.
The results are shown in Figure~\ref{fig:fitbe1}. 
We checked that for all $\rho/\mu_e$ (only eight values are shown in Figure~\ref{fig:fitbe1}), the deviation between the tabulated rate and the fit is less than 25~\%.

%-----------Table --------------------
\begin{table}[H]
\scriptsize{
\caption{Fit parameters of the $^{7}$Be($e^-,\nu_e$)$^{7}$Li decay rate.} 
\label{table:fits}

\resizebox{13.86cm}{!} {
\begin{tabular}{lccccc} 
\toprule
   & \boldmath{$b_{k,0}$} & \boldmath{$b_{k,1}$}  & \boldmath{$b_{k,2}$} & \boldmath{$b_{k,3}$}  & \boldmath{$b_{k,4}$} \\ 
\midrule
$k=0$   & $2.501 $ & $-9.756 \times 10^{-1}$ &  $1.937 \times 10^{-2}$ &  $1.262 \times 10^{-2}$ &  $2.041 \times 10^{-3}$ \\  
$k=1$   & $6.027 $ &  $2.897 \times 10^{-1}$ &  $6.012 \times 10^{-2}$ &  $6.633 \times 10^{-3}$ & $-$ \\ 
$k=2$   & $1.808 \times 10^{-1}$  &  $4.836 \times 10^{-2}$ & $-8.658 \times 10^{-4}$ & $-3.794 \times 10^{-3}$ & $-5.807 \times 10^{-4}$ \\ 
$k=3$   &$9.491 \times 10^{-1}$ & $2.574 \times 10^{-2}$ & $-$ & $-$ & $-$ \\ 
$k=4$   &$2.308 \times 10^{-1}$ &  $6.432 \times 10^{-2}$ & $9.362 \times 10^{-3}$  & $-2.524 \times 10^{-4}$ & $-1.669 \times 10^{-4}$ \\ 
$k=5$   &$-9.343 $ & $1.011 $ &$4.900 \times 10^{-3}$ & $-$ & $-$ \\ 
\bottomrule
\end{tabular}
}

}
\end{table}

%%%%%%%%%%%%%%%%%%%%%%%%%%%%%%%%%%%%%%%%%%
\begin{adjustwidth}{-\extralength}{0cm}
\printendnotes[custom] % Un-comment to print a list of endnotes

\reftitle{References}

% Please provide either the correct journal abbreviation (e.g., according to the “List of Title Word Abbreviations” http://www.issn.org/services/online-services/access-to-the-ltwa/) or the full name of the journal.


%-----------BIBLIO--------------

%\bibliography{astro.bib}

%----------------------------------------------------

\PublishersNote{}
\end{adjustwidth}
\end{document}